\documentclass[twocolumn, aps, prd, superscriptaddress,floatfix]{revtex4-1}

\usepackage{multirow}
\usepackage{amsmath,amssymb}
\usepackage{graphicx,bm}
\usepackage{slashed}
\usepackage{epstopdf}
\usepackage{ulem} 
\usepackage[usenames]{color}
\usepackage{float}
\usepackage{braket}
\usepackage{appendix}
\usepackage{hyperref}
\usepackage{ytableau}

\renewcommand\sout{\bgroup \color{red} \ULdepth=-.5ex \ULset}

\begin{document}

\title{Exploring bound states and interactions of the nucleon–antinucleon system in a constituent quark model}

\author{Aaron Park}
\email{aaron.park@yonsei.ac.kr}\affiliation{Department of Physics and Institute of Physics and Applied Physics, Yonsei University, Seoul 03722, Korea}

\author{Sungtae Cho}\email{sungtae.cho@kangwon.ac.kr}
\affiliation{Division of Science Education, Kangwon National University, Chuncheon 24341, Korea}

\author{Su Houng Lee}%
\email{suhoung@yonsei.ac.kr}
\affiliation{Department of Physics and Institute of Physics and Applied Physics, Yonsei University, Seoul 03722, Korea}

\date{\today}
\begin{abstract}
In this work, we study the nucleon-antinucleon  system in a constituent quark model. We first construct the nucleon–antinucleon wave function such that the multiquark  and multiantiquark components each satisfy the Pauli exclusion principle, and then investigate the possibility of a bound state using a Hamiltonian that includes color–color, spin-dependent interactions and three-quark potentials. Our results indicate that, in specific channels, the nucleon–antinucleon interaction exhibits significant attraction, suggesting a strong possibility of a bound state.
\end{abstract}

\maketitle

\section{Introduction}
\label{Introduction}

The study of multiquark systems has long been a subject of interest in hadronic physics, offering insights into the quantum chromodynamics(QCD) beyond the conventional quark–antiquark and three-quark configurations \cite{Jaffe:1976ig,Jaffe:1976yi,Swanson:2006st,Nielsen:2009uh,Richard:2016eis}. According to color SU(3), a color singlet state must be formed from a combination of quarks and antiquarks in the form of $q^{n+3m} \overline{q}^{n}$, where $n$ and $m$ are integers. Based on this criterion, not only tetraquarks, pentaquarks, and dibaryons, but also configurations consisting of a baryon and an antibaryon, such as the $q^3 \overline{q}^3$ system, can form a color singlet state. 

Although the nucleon and antinucleon are well-understood as individual three-quark and three-antiquark systems, respectively, the composite nucleon-antinucleon systems remain an open theoretical question \cite{Shapiro:1978wi,Dover:1992vj,Klempt:2002ap,Klempt:2005pp,Ortega:2024zjx,Entem:2006dt}. The baryon–antibaryon configurations can provide a particularly compelling framework for investigating the role of color confinement and spin-dependent interactions at the multiquark level. Additionally, the possibility of forming a tightly bound or resonant state in the $N\overline{N}$ system has implications not only for hadron spectroscopy but also for understanding short-range nucleon–antinucleon interactions, which may play a role in annihilation processes and exotic state formation.

There are several experimental observations that indicate possible $p \bar{p}$ bound states.  BESIII observed threshold  enhancement of $ p \bar{p}$ in the $J/\psi \rightarrow \gamma p \bar{p}$ decay\cite{BES:2003aic}
and a resonance peak $X(1835)$ was seen in the $J/\psi \rightarrow \pi^+ \pi^- \eta'$ decay\cite{BES:2005ega}. Recently, anomalous peak structure of X(1840) in the $J/\psi \rightarrow \gamma 3(\pi^+ \pi^-)$ was observed, which could be well explained with overlapping resonant structures X(1840) and X(1880)\cite{BESIII:2023vvr}. 

Previous theoretical studies have approached the $p\bar{p}$ problem using various frameworks, including potential models and effective field theories \cite{Datta:2003iy,Chang:2004us,Yan:2004xs,Wang:2006sna,Kang:2013uia}.

However, many of these approaches either treat the $N\overline{N}$ system as a molecular state or neglect the full multiquark structure and the symmetrization constraints imposed by the Pauli exclusion principle. As has been shown in previous studies, the $N\overline{N}$ system exhibits more attractive interaction than the nucleon–nucleon system due to G-parity effect \cite{Klempt:2002ap}. Therefore, the $N\overline{N}$ system is expected to have a high possibility of forming a bound state with a binding energy larger than that of the deuteron. However, previous studies have primarily examined the interaction from the perspective of intermediate and long-range potentials, while the short-range interaction has not been adequately investigated.  For this reason, it should be studied not merely as an interaction between two hadrons, but rather as a compact six-quark state composed of three quarks and three antiquarks.
From this point of view, the constituent quark model can provide a useful platform to construct an antisymmetrized wave functions and analyze multiquark interactions through color–spin-dependent forces.

In this work, we investigate the nucleon-antinucleon system within the constituent quark model by explicitly constructing a six-quark wave function in which both the nucleon and antinucleon components satisfy the Pauli exclusion principle independently. Using a Hamiltonian that includes both color–color, color–spin, and three-quark interactions we analyze the possibility of a compact bound state. Our findings indicate that a bound state may indeed emerge when the total spin of the system is zero, suggesting a potential mechanism for binding in the $N\overline{N}$ channel that is not purely molecular in nature.

This paper is organized as follows. In Sec.\ref{Hamiltonian and Jacobi coordinate}, we introduce the Hamiltonian and appropriate coordinate system to describe the baryon-antibaryon system. In Sec.\ref{Color basis}, we present the color basis of $q^3 \overline{q}^3$ using Young-Yamanouchi basis. In Sec.\ref{flavor-color-spin}, we construct the wave function of nucleon-antinucleon satisfying the Pauli exclusion principle. In Sec.\ref{Results}, we investigate the interaction between nucleon and antinucleon using $q^3 \overline{q}^3$ configuration and the possibility of compact nucleon-antinucleon state. Sec.\ref{Summary} is devoted to summary and concluding remarks.

\section{Hamiltonian and Jacobi coordinate}
\label{Hamiltonian and Jacobi coordinate}

In this work, we use the following quark model Hamiltonian \cite{Noh:2024nvp,Park:2024tox}. 
\begin{eqnarray}
H &=& \sum^{n}_{i=1} \left( m_i+\frac{{\mathbf p}^{2}_i}{2 m_i} \right)-\frac{3}{4}\sum^{n}_{i<j}\frac{\lambda^{c}_{i}}{2} \,\, \frac{\lambda^{c}_{j}}{2} \left( V^{C}_{ij} + V^{CS}_{ij} \right) \nonumber \\
&& + \sum^n_{i<j<k} \left( AL^{C-C}_{ijk} + BL^{S-S}_{ijk} + CL^{C-S}_{ijk} \right),
\label{Hamiltonian}
\end{eqnarray}
where $n$ is the number of quarks, $m_i$'s are the constituent quark masses, and $\lambda_i^c$ are the Gell-Mann matrices of the $i$'th quark for the color SU(3). The two-body quark potentials which are composed of color-color interaction  $V^C_{ij}$ and color-spin interaction  $V^{CS}_{ij}$ are as follows.
\begin{eqnarray}
V^{C}_{ij} &=& - \frac{\kappa}{r_{ij}} + \frac{r_{ij}}{a^2_0} - D,
\label{ConfineP}
\\
V^{CS}_{ij} &=& \frac{\hbar^2 c^2 \kappa'}{m_i m_j c^4} \frac{e^{- \left( r_{ij} \right)^2 / \left( r_{0ij} \right)^2}}{(r_{0ij}) r_{ij}} \sigma_i \cdot \sigma_j\,.
\label{CSP}
\end{eqnarray}
Here, $r_{0 ij} = \left( \alpha + \beta m_{ij}  \right)^{-1}$, $\kappa' = \kappa_0 \left( 1 + \gamma m_{ij} \right)$ and $m_{ij}=\frac{m_i m_j}{m_i + m_j}$. The parameters are as follows.
$\kappa=87.86 \, \textrm{MeV fm}$, $a_0=0.02993 \, \textrm{(MeV$^{-1}$fm)$^{1/2}$}$, $D=1082  \, \textrm{MeV}$, $ m_{u}=m_{d}=315 \, \textrm{MeV}$, $\alpha = 1.0338 \, \textrm{fm$^{-1}$}$, $\beta = 0.0005885 \, \textrm{(MeV fm)$^{-1}$}$, $\gamma = 0.0010444 \, \textrm{MeV$^{-1}$}$, $\kappa_0=220.06 \, \textrm{MeV}$. The detailed form of the three-quark potential $L^{C-C}$, $L^{S=S}$, and $L^{C-S}$ are provided in Ref \cite{Noh:2024nvp,Park:2024tox}. The three-quark parameters are $A = -685 \, \textrm{MeV}^2$, $B = 6.43 \times 10^{11} \, \textrm{MeV}^6$, $C = -1.24 \times 10^7 \, \textrm{MeV}^4$. The parameters are determined by fitting to the $N$, $\Delta$, $\Lambda$, $\Sigma$, and $\Sigma^*$ baryon spectra. 
The fit results are presented in Table
 \ref{baryons}. The two variational parameters $a_1$ and $a_2$ shown in the Table are parameters of the spatial Gaussian wave function and are associated with the interquark distances.

\begin{table}[ht]
\centering
\begin{tabular}{cccc}
\hline
\hline	\multirow{2}{*}{Particle}	&	Experimental 	&	Mass		&	\quad Variational\quad					\\
& Value (MeV)	&	(MeV)	&	\quad Parameters (${\rm fm}^{-2}$)\quad	\\
\hline
$p$				&	938.27	&	939.2 (1007.0)	&	\quad$a_1$ = 2.4, $a_2$ = 2.4\quad	\\
$\Delta$		&	1232	&	1231.4 (1349.2)	&	\quad$a_1$ = 1.8, $a_2$ = 1.8\quad	\\
$\Lambda$		&	1115.7	&	1096.4 (1141.9)	&	\quad$a_1$ = 2.8, $a_2$ = 2.7\quad	\\
$\Sigma$		&	1192.6	&	1164.8 (1241.9)	&	\quad$a_1$ = 2.1, $a_2$ = 3.1\quad	\\
$\Sigma^*$&	1383.7	&	1369.6 (1464.4)	&	\quad$a_1$ = 1.9, $a_2$ = 2.4\quad	\\
\hline 
\hline
\label{baryons}
\end{tabular}
\caption{Baryon fitting results. Values in parentheses correspond to fits obtained without the three-quark potentials.}
\end{table}

In order to formulate the spatial component of the nucleon–antinucleon wave function, we utilize the following Jacobi coordinate system.
\begin{align}
    \vec{x}_1 &= \frac{1}{\sqrt{2}}(\vec{r}_1-\vec{r}_2), \\
    \vec{x}_2 &= \frac{1}{\sqrt{6}}(\vec{r}_1 + \vec{r}_2 -2\vec{r}_3 ), \\
    \vec{x}_3 &= \frac{1}{\sqrt{2}}(\vec{r}_4-\vec{r}_5), \\
    \vec{x}_4 &= \frac{1}{\sqrt{6}}(\vec{r}_4 + \vec{r}_5 -2\vec{r}_6 ), \\
    \vec{x}_5 &= \frac{1}{\sqrt{6}}(\vec{r}_1 + \vec{r}_2 + \vec{r}_3 -\vec{r}_4 -\vec{r}_5 -\vec{r}_6 ). 
\end{align}
In this work, the spatial part of the wave function is taken to be of the following Gaussian form.
\begin{align}
    R=e^{-a_1 x_1^2-a_1 x_2^2-a_2 x_3^2-a_2 x_4^2-a_3 x_5^2}
\end{align}

\section{Color basis}
\label{Color basis}

There are two main approaches to obatin the color singlet basis of $q^3 \overline{q}^3$: baryon-antibaryon configuration and three-meson configuration. We begin by ordering the quarks as in $q(1)q(2)q(3)\overline{q}(4) \overline{q}(5) \overline{q}(6)$ and first consider the three-meson configuration. In color SU(3), a quark transforms as a triplet and an antiquark transforms as an antitriplet. Therefore, the product of three quarks and three antiquarks can be expressed as follows.
\begin{align}
    & (\mathbf{3}_1 \times \overline{\mathbf{3}}_4) \times (\mathbf{3}_2 \times \overline{\mathbf{3}}_5) \times (\mathbf{3}_3 \times \overline{\mathbf{3}}_6) \nonumber \\
    &= (\mathbf{1}+\mathbf{8})_{14} \times (\mathbf{1}+\mathbf{8})_{25} \times (\mathbf{1}+\mathbf{8})_{36} \nonumber \\
    &= [\mathbf{1}(\mathbf{1}_{14}\mathbf{1}_{25}) + \mathbf{8}(\mathbf{1}_{14}\mathbf{8}_{25}) + \mathbf{8}(\mathbf{8}_{14}\mathbf{1}_{25}) \nonumber + \mathbf{1}(\mathbf{8}_{14}\mathbf{8}_{25}) \nonumber \\
    & + \mathbf{8}(\mathbf{8}_{14}\mathbf{8}_{25}) + \mathbf{8}'(\mathbf{8}_{14}\mathbf{8}_{25}) + \mathbf{10}(\mathbf{8}_{14}\mathbf{8}_{25}) + \overline{\mathbf{10}}(\mathbf{8}_{14}\mathbf{8}_{25}) \nonumber \\
    &+ \mathbf{27}(\mathbf{8}_{14}\mathbf{8}_{25})] \times (\mathbf{1}+\mathbf{8})_{36},
\end{align}
where the subscripts denote the quark labels.
So, there are six color singlet states as follows.\\

$\mathbf{1}_{14}\mathbf{1}_{25}\mathbf{1}_{36}$, $\mathbf{1}_{14}\mathbf{8}_{25}\mathbf{8}_{36}$, $\mathbf{8}_{14}\mathbf{1}_{25}\mathbf{8}_{36}$, $\mathbf{8}_{14}\mathbf{8}_{25}\mathbf{1}_{36}$, \\

$\mathbf{8}_{14}\mathbf{8}_{25}\mathbf{8}_{36}$, $(\mathbf{8}_{14}\mathbf{8}_{25}\mathbf{8}_{36})'$.\\

We now turn to constructing the color basis in the baryon–antibaryon configuration.
\begin{align}
    &\mathbf{3}_1 \times \mathbf{3}_2 \times \mathbf{3}_3 \times \overline{\mathbf{3}}_4 \times \overline{\mathbf{3}}_5 \times \overline{\mathbf{3}}_6 \nonumber \\
    &= (\mathbf{1}_{123}+\mathbf{8}_{123}+\mathbf{8}_{123}+\mathbf{10}_{123})\nonumber \\
    & \quad \times (\mathbf{1}_{456}+\mathbf{8}_{456}+\mathbf{8}_{456}+\overline{\mathbf{10}}_{456})
\end{align}
In this case as well, there are six color singlet states: one $\mathbf{1}_{123}\mathbf{1}_{456}$, four $\mathbf{8}_{123}\mathbf{8}_{456}$ and one $\mathbf{10}_{123}\overline{\mathbf{10}}_{456}$. If we construct the color singlet basis in this way, the quark part and the antiquark part share the same structure, so we can obtain the color singlet simply by contracting the corresponding indices. Additionally, we can represent these color basis using the Young-Yamanouchi symbol. For simplicity, the ket notation is omitted throughout.\\

$C_1 = \left( \begin{tabular}{|c|}
  \hline
  $\hspace{0.08cm} 1\hspace{0.08cm}$ \\
  \hline
  $\hspace{0.08cm} 2\hspace{0.08cm}$ \\
  \hline
  $\hspace{0.08cm} 3\hspace{0.08cm}$ \\
  \hline
\end{tabular},
\begin{tabular}{|c|}
  \hline
  $\hspace{0.08cm} \overline{4}\hspace{0.08cm}$ \\
  \hline
  $\hspace{0.08cm} \overline{5}\hspace{0.08cm}$ \\
  \hline
  $\hspace{0.08cm} \overline{6}\hspace{0.08cm}$ \\
  \hline
\end{tabular} \right)$,
$C_2 = \left( \begin{tabular}{|c|c|}
  \hline
  $\hspace{0.08cm} 1\hspace{0.08cm}$ & $\hspace{0.08cm} 2\hspace{0.08cm}$ \\
  \hline
  $\hspace{0.08cm} 3\hspace{0.08cm}$ \\
  \cline{1-1}
\end{tabular},
\begin{tabular}{|c|c|}
  \hline
  $\hspace{0.08cm} \overline{4}\hspace{0.08cm}$ & $\hspace{0.08cm} \overline{5}\hspace{0.08cm}$ \\
  \hline
  $\hspace{0.08cm} \overline{6}\hspace{0.08cm}$ \\
  \cline{1-1}
\end{tabular} \right)$,\\

$C_3 = \left( \begin{tabular}{|c|c|}
  \hline
  $\hspace{0.08cm} 1\hspace{0.08cm}$ & $\hspace{0.08cm} 2\hspace{0.08cm}$ \\
  \hline
  $\hspace{0.08cm} 3\hspace{0.08cm}$ \\
  \cline{1-1}
\end{tabular},
\begin{tabular}{|c|c|}
  \hline
  $\hspace{0.08cm} \overline{4}\hspace{0.08cm}$ & $\hspace{0.08cm} \overline{6}\hspace{0.08cm}$ \\
  \hline
  $\hspace{0.08cm} \overline{5}\hspace{0.08cm}$ \\
  \cline{1-1}
\end{tabular} \right)$,
$C_4 = \left( \begin{tabular}{|c|c|}
  \hline
  $\hspace{0.08cm} 1\hspace{0.08cm}$ & $\hspace{0.08cm} 3\hspace{0.08cm}$ \\
  \hline
  $\hspace{0.08cm} 2\hspace{0.08cm}$ \\
  \cline{1-1}
\end{tabular},
\begin{tabular}{|c|c|}
  \hline
  $\hspace{0.08cm} \overline{4}\hspace{0.08cm}$ & $\hspace{0.08cm} \overline{5}\hspace{0.08cm}$ \\
  \hline
  $\hspace{0.08cm} \overline{6}\hspace{0.08cm}$ \\
  \cline{1-1}
\end{tabular} \right)$,\\

$C_5 = \left( \begin{tabular}{|c|c|}
  \hline
  $\hspace{0.08cm} 1\hspace{0.08cm}$ & $\hspace{0.08cm} 3\hspace{0.08cm}$ \\
  \hline
  $\hspace{0.08cm} 2\hspace{0.08cm}$ \\
  \cline{1-1}
\end{tabular},
\begin{tabular}{|c|c|}
  \hline
  $\hspace{0.08cm} \overline{4}\hspace{0.08cm}$ & $\hspace{0.08cm} \overline{6}\hspace{0.08cm}$ \\
  \hline
  $\hspace{0.08cm} \overline{5}\hspace{0.08cm}$ \\
  \cline{1-1}
\end{tabular} \right)$,
$C_6 = \left( \begin{tabular}{|c|c|c|}
  \hline
  $\hspace{0.08cm} 1\hspace{0.08cm}$ & $\hspace{0.08cm} 2\hspace{0.08cm}$ & $\hspace{0.08cm} 3\hspace{0.08cm}$ \\
  \hline
\end{tabular},
\begin{tabular}{|c|c|c|}
  \hline
  $\hspace{0.08cm} \overline{4}\hspace{0.08cm}$ & $\hspace{0.08cm} \overline{5}\hspace{0.08cm}$ & $\hspace{0.08cm} \overline{6}\hspace{0.08cm}$ \\
  \hline
\end{tabular} \right)$\\

In this work, we consider only baryon-antibaryon configuration and leave the three-meson configuration as future work. We represent the color-color matrices for this baryon-antibaryon configuration in the appendix \ref{Color-color matrices}.

\section{Flavor-color-spin state of $q^3 \overline{q}^3$}
\label{flavor-color-spin}

In this section, we construct the wave function of $q^3 \overline{q}^3$, where $q=u,d$, to satisfy the Pauli exclusion principle. Since we are interested in the lowest-energy configuration, we assume that the spatial part of the wave function to be symmetric. 
Therefore, the remaining part of the wave function which is the flavor-color-spin part should be antisymmetric for quarks and antiquarks, respectively.

\subsection{$I=1$}

In order for the total isospin of the $q^3 \overline{q}^3$ system to be 1, the possible isospin combinations of the quarks and antiquarks are ($\frac{3}{2}$,$\frac{3}{2}$),($\frac{3}{2}$,$\frac{1}{2}$),($\frac{1}{2}$,$\frac{3}{2}$), and ($\frac{1}{2}$,$\frac{1}{2}$). For the wave function to be antisymmetric, the color-spin part must be the conjugate of the flavor part. We represent the flavor $(F)$ and color-spin basis $(CS) $ of $q^3 \overline{q}^3$ with $I=1$ as follows.

$F_1^A = (\begin{tabular}{|c|c|c|}
   \hline
   $\hspace{0.08cm} 1\hspace{0.08cm}$ & $\hspace{0.08cm} 2\hspace{0.08cm}$ & $\hspace{0.08cm} 3\hspace{0.08cm}$  \\
   \hline
\end{tabular},
\begin{tabular}{|c|c|c|}
    \hline
    $\hspace{0.08cm} \overline{4}\hspace{0.08cm}$ & $\hspace{0.08cm} \overline{5}\hspace{0.08cm}$ & $\hspace{0.08cm} \overline{6}\hspace{0.08cm}$ \\ 
    \hline
\end{tabular}),\
CS_1^A = \left(
\begin{tabular}{|c|}
    \hline
    $\hspace{0.08cm} 1\hspace{0.08cm}$ \\
    \hline
    $\hspace{0.08cm} 2\hspace{0.08cm}$ \\
    \hline
    $\hspace{0.08cm} 3\hspace{0.08cm}$ \\
    \hline
\end{tabular},
\begin{tabular}{|c|}
    \hline
    $\hspace{0.08cm} \overline{4}\hspace{0.08cm}$ \\
    \hline
    $\hspace{0.08cm} \overline{5}\hspace{0.08cm}$ \\
    \hline
    $\hspace{0.08cm} \overline{6}\hspace{0.08cm}$ \\
    \hline
\end{tabular} \right),
$

$F_1^B = (\begin{tabular}{|c|c|c|}
   \hline
   $\hspace{0.08cm} 1\hspace{0.08cm}$ & $\hspace{0.08cm} 2\hspace{0.08cm}$ & $\hspace{0.08cm} 3\hspace{0.08cm}$  \\
   \hline
\end{tabular},
\begin{tabular}{|c|c|}
    \hline
    $\hspace{0.08cm} \overline{4}\hspace{0.08cm}$ & $\hspace{0.08cm} \overline{5}\hspace{0.08cm}$ \\
    \hline
    $\hspace{0.08cm} \overline{6}\hspace{0.08cm}$ \\ 
    \cline{1-1}
\end{tabular}),\
F_2^B = (\begin{tabular}{|c|c|c|}
   \hline
   $\hspace{0.08cm} 1\hspace{0.08cm}$ & $\hspace{0.08cm} 2\hspace{0.08cm}$ & $\hspace{0.08cm} 3\hspace{0.08cm}$  \\
   \hline
\end{tabular},
\begin{tabular}{|c|c|}
    \hline
    $\hspace{0.08cm} \overline{4}\hspace{0.08cm}$ & $\hspace{0.08cm} \overline{6}\hspace{0.08cm}$ \\
    \hline
    $\hspace{0.08cm} \overline{5}\hspace{0.08cm}$ \\ 
    \cline{1-1}
\end{tabular}),$\\

$CS_1^B = \left(
\begin{tabular}{|c|}
    \hline
    $\hspace{0.08cm} 1\hspace{0.08cm}$ \\
    \hline
    $\hspace{0.08cm} 2\hspace{0.08cm}$ \\
    \hline
    $\hspace{0.08cm} 3\hspace{0.08cm}$ \\
    \hline
\end{tabular},
\begin{tabular}{|c|c|}
    \hline
    $\hspace{0.08cm} \overline{4}\hspace{0.08cm}$ &
    $\hspace{0.08cm} \overline{5}\hspace{0.08cm}$ \\
    \hline
    $\hspace{0.08cm} \overline{6}\hspace{0.08cm}$ \\
    \cline{1-1}
\end{tabular} \right), \
CS_2^B = \left(
\begin{tabular}{|c|}
    \hline
    $\hspace{0.08cm} 1\hspace{0.08cm}$ \\
    \hline
    $\hspace{0.08cm} 2\hspace{0.08cm}$ \\
    \hline
    $\hspace{0.08cm} 3\hspace{0.08cm}$ \\
    \hline
\end{tabular},
\begin{tabular}{|c|c|}
    \hline
    $\hspace{0.08cm} \overline{4}\hspace{0.08cm}$ &
    $\hspace{0.08cm} \overline{6}\hspace{0.08cm}$ \\
    \hline
    $\hspace{0.08cm} \overline{5}\hspace{0.08cm}$ \\
    \cline{1-1}
\end{tabular} \right),
$\\

$F_1^C = (\begin{tabular}{|c|c|c|}
   \hline
   $\hspace{0.08cm} 1\hspace{0.08cm}$ & $\hspace{0.08cm} 2\hspace{0.08cm}$ & $\hspace{0.08cm} 3\hspace{0.08cm}$  \\
   \hline
\end{tabular},
\begin{tabular}{|c|c|}
    \hline
    $\hspace{0.08cm} \overline{4}\hspace{0.08cm}$ & $\hspace{0.08cm} \overline{5}\hspace{0.08cm}$ \\
    \hline
    $\hspace{0.08cm} \overline{6}\hspace{0.08cm}$ \\ 
    \cline{1-1}
\end{tabular}),\
F_2^C = (\begin{tabular}{|c|c|c|}
   \hline
   $\hspace{0.08cm} 1\hspace{0.08cm}$ & $\hspace{0.08cm} 2\hspace{0.08cm}$ & $\hspace{0.08cm} 3\hspace{0.08cm}$  \\
   \hline
\end{tabular},
\begin{tabular}{|c|c|}
    \hline
    $\hspace{0.08cm} \overline{4}\hspace{0.08cm}$ & $\hspace{0.08cm} \overline{6}\hspace{0.08cm}$ \\
    \hline
    $\hspace{0.08cm} \overline{5}\hspace{0.08cm}$ \\ 
    \cline{1-1}
\end{tabular})$,\\

$CS_1^C = \left(
\begin{tabular}{|c|}
    \hline
    $\hspace{0.08cm} 1\hspace{0.08cm}$ \\
    \hline
    $\hspace{0.08cm} 2\hspace{0.08cm}$ \\
    \hline
    $\hspace{0.08cm} 3\hspace{0.08cm}$ \\
    \hline
\end{tabular},
\begin{tabular}{|c|c|}
    \hline
    $\hspace{0.08cm} \overline{4}\hspace{0.08cm}$ &
    $\hspace{0.08cm} \overline{5}\hspace{0.08cm}$ \\
    \hline
    $\hspace{0.08cm} \overline{6}\hspace{0.08cm}$ \\
    \cline{1-1}
\end{tabular} \right), \
CS_2^C = \left(
\begin{tabular}{|c|}
    \hline
    $\hspace{0.08cm} 1\hspace{0.08cm}$ \\
    \hline
    $\hspace{0.08cm} 2\hspace{0.08cm}$ \\
    \hline
    $\hspace{0.08cm} 3\hspace{0.08cm}$ \\
    \hline
\end{tabular},
\begin{tabular}{|c|c|}
    \hline
    $\hspace{0.08cm} \overline{4}\hspace{0.08cm}$ &
    $\hspace{0.08cm} \overline{6}\hspace{0.08cm}$ \\
    \hline
    $\hspace{0.08cm} \overline{5}\hspace{0.08cm}$ \\
    \cline{1-1}
\end{tabular} \right),
$\\

$F_1^D = (\begin{tabular}{|c|c|}
   \hline
   $\hspace{0.08cm} 1\hspace{0.08cm}$ & $\hspace{0.08cm} 2\hspace{0.08cm}$ \\
   \hline
   $\hspace{0.08cm} 3\hspace{0.08cm}$  \\
   \cline{1-1}
\end{tabular},
\begin{tabular}{|c|c|}
    \hline
    $\hspace{0.08cm} \overline{4}\hspace{0.08cm}$ & $\hspace{0.08cm} \overline{5}\hspace{0.08cm}$ \\
    \hline
    $\hspace{0.08cm} \overline{6}\hspace{0.08cm}$ \\ 
    \cline{1-1}
\end{tabular}),\
F_2^D = (\begin{tabular}{|c|c|}
   \hline
   $\hspace{0.08cm} 1\hspace{0.08cm}$ & $\hspace{0.08cm} 2\hspace{0.08cm}$ \\
   \hline
   $\hspace{0.08cm} 3\hspace{0.08cm}$  \\
   \cline{1-1}
\end{tabular},
\begin{tabular}{|c|c|}
    \hline
    $\hspace{0.08cm} \overline{4}\hspace{0.08cm}$ & $\hspace{0.08cm} \overline{5}\hspace{0.08cm}$ \\
    \hline
    $\hspace{0.08cm} \overline{6}\hspace{0.08cm}$ \\ 
    \cline{1-1}
\end{tabular})$,\\

$F_3^D = (\begin{tabular}{|c|c|}
   \hline
   $\hspace{0.08cm} 1\hspace{0.08cm}$ & $\hspace{0.08cm} 3\hspace{0.08cm}$ \\
   \hline
   $\hspace{0.08cm} 2\hspace{0.08cm}$  \\
   \cline{1-1}
\end{tabular},
\begin{tabular}{|c|c|}
    \hline
    $\hspace{0.08cm} \overline{4}\hspace{0.08cm}$ & $\hspace{0.08cm} \overline{5}\hspace{0.08cm}$ \\
    \hline
    $\hspace{0.08cm} \overline{6}\hspace{0.08cm}$ \\ 
    \cline{1-1}
\end{tabular}),\
F_4^D = (\begin{tabular}{|c|c|}
   \hline
   $\hspace{0.08cm} 1\hspace{0.08cm}$ & $\hspace{0.08cm} 3\hspace{0.08cm}$ \\
   \hline
   $\hspace{0.08cm} 2\hspace{0.08cm}$  \\
   \cline{1-1}
\end{tabular},
\begin{tabular}{|c|c|}
    \hline
    $\hspace{0.08cm} \overline{4}\hspace{0.08cm}$ & $\hspace{0.08cm} \overline{6}\hspace{0.08cm}$ \\
    \hline
    $\hspace{0.08cm} \overline{5}\hspace{0.08cm}$ \\ 
    \cline{1-1}
\end{tabular}),$ \\

$CS_1^D = (\begin{tabular}{|c|c|}
   \hline
   $\hspace{0.08cm} 1\hspace{0.08cm}$ & $\hspace{0.08cm} 2\hspace{0.08cm}$ \\
   \hline
   $\hspace{0.08cm} 3\hspace{0.08cm}$  \\
   \cline{1-1}
\end{tabular},
\begin{tabular}{|c|c|}
    \hline
    $\hspace{0.08cm} \overline{4}\hspace{0.08cm}$ & $\hspace{0.08cm} \overline{5}\hspace{0.08cm}$ \\
    \hline
    $\hspace{0.08cm} \overline{6}\hspace{0.08cm}$ \\ 
    \cline{1-1}
\end{tabular}),\
CS_2^D = (\begin{tabular}{|c|c|}
   \hline
   $\hspace{0.08cm} 1\hspace{0.08cm}$ & $\hspace{0.08cm} 2\hspace{0.08cm}$ \\
   \hline
   $\hspace{0.08cm} 3\hspace{0.08cm}$  \\
   \cline{1-1}
\end{tabular},
\begin{tabular}{|c|c|}
    \hline
    $\hspace{0.08cm} \overline{4}\hspace{0.08cm}$ & $\hspace{0.08cm} \overline{6}\hspace{0.08cm}$ \\
    \hline
    $\hspace{0.08cm} \overline{5}\hspace{0.08cm}$ \\ 
    \cline{1-1}
\end{tabular})$,\\

$CS_3^D = (\begin{tabular}{|c|c|}
   \hline
   $\hspace{0.08cm} 1\hspace{0.08cm}$ & $\hspace{0.08cm} 3\hspace{0.08cm}$ \\
   \hline
   $\hspace{0.08cm} 2\hspace{0.08cm}$  \\
   \cline{1-1}
\end{tabular},
\begin{tabular}{|c|c|}
    \hline
    $\hspace{0.08cm} \overline{4}\hspace{0.08cm}$ & $\hspace{0.08cm} \overline{5}\hspace{0.08cm}$ \\
    \hline
    $\hspace{0.08cm} \overline{6}\hspace{0.08cm}$ \\ 
    \cline{1-1}
\end{tabular}),\
CS_4^D = (\begin{tabular}{|c|c|}
   \hline
   $\hspace{0.08cm} 1\hspace{0.08cm}$ & $\hspace{0.08cm} 3\hspace{0.08cm}$ \\
   \hline
   $\hspace{0.08cm} 2\hspace{0.08cm}$  \\
   \cline{1-1}
\end{tabular},
\begin{tabular}{|c|c|}
    \hline
    $\hspace{0.08cm} \overline{4}\hspace{0.08cm}$ & $\hspace{0.08cm} \overline{6}\hspace{0.08cm}$ \\
    \hline
    $\hspace{0.08cm} \overline{5}\hspace{0.08cm}$ \\ 
    \cline{1-1}
\end{tabular}).$\\
Here, the superscripts $A,B,C,$ and $D$ correspond to the isospin ($\frac{3}{2}$,$\frac{3}{2}$),($\frac{3}{2}$,$\frac{1}{2}$),($\frac{1}{2}$,$\frac{3}{2}$), and ($\frac{1}{2}$,$\frac{1}{2}$), respectively. In addition, for the three-quark system with isospin $\frac{1}{2}$, there exist two independent bases, which are indicated by subscripts. In order to construct a totally antisymmetric wave function among the quarks, as well as among the antiquarks, all of these basis set must be taken into account. Note that in the $D$ case, both the three-quark and three-antiquark systems have isospin $\frac{1}{2}$, resulting in four independent basis states.

Finally, the color-spin coupling state can be expressed in terms of the color and spin basis functions using Clebsch–Gordan coefficients. For the spin basis, we can use the same one as dibaryon \cite{Park:2016mez}. The explicit forms of the flavor–color–spin wave functions of $q^3 \overline{q}^3$ are presented in Appendix~\ref{wave-function}.

\subsection{$I=0$}

In order for the total isospin of the $q^3 \overline{q}^3$ system to be 0, the possible isospin combinations of the quarks and antiquarks are ($\frac{3}{2}$,$\frac{3}{2}$) and ($\frac{1}{2}$,$\frac{1}{2}$). We represent the flavor and color-spin basis of $q^3 \overline{q}^3$ with $I=0$ as follows.

$F_1^A = (\begin{tabular}{|c|c|c|}
   \hline
   $\hspace{0.08cm} 1\hspace{0.08cm}$ & $\hspace{0.08cm} 2\hspace{0.08cm}$ & $\hspace{0.08cm} 3\hspace{0.08cm}$  \\
   \hline
\end{tabular},
\begin{tabular}{|c|c|c|}
    \hline
    $\hspace{0.08cm} \overline{4}\hspace{0.08cm}$ & $\hspace{0.08cm} \overline{5}\hspace{0.08cm}$ & $\hspace{0.08cm} \overline{6}\hspace{0.08cm}$ \\ 
    \hline
\end{tabular}),\
CS_1^A = \left(
\begin{tabular}{|c|}
    \hline
    $\hspace{0.08cm} 1\hspace{0.08cm}$ \\
    \hline
    $\hspace{0.08cm} 2\hspace{0.08cm}$ \\
    \hline
    $\hspace{0.08cm} 3\hspace{0.08cm}$ \\
    \hline
\end{tabular},
\begin{tabular}{|c|}
    \hline
    $\hspace{0.08cm} \overline{4}\hspace{0.08cm}$ \\
    \hline
    $\hspace{0.08cm} \overline{5}\hspace{0.08cm}$ \\
    \hline
    $\hspace{0.08cm} \overline{6}\hspace{0.08cm}$ \\
    \hline
\end{tabular} \right)
$

$F_1^B = (\begin{tabular}{|c|c|}
   \hline
   $\hspace{0.08cm} 1\hspace{0.08cm}$ & $\hspace{0.08cm} 2\hspace{0.08cm}$ \\
   \hline
   $\hspace{0.08cm} 3\hspace{0.08cm}$  \\
   \cline{1-1}
\end{tabular},
\begin{tabular}{|c|c|}
    \hline
    $\hspace{0.08cm} \overline{4}\hspace{0.08cm}$ & $\hspace{0.08cm} \overline{5}\hspace{0.08cm}$ \\
    \hline
    $\hspace{0.08cm} \overline{6}\hspace{0.08cm}$ \\ 
    \cline{1-1}
\end{tabular}),\
F_2^B = (\begin{tabular}{|c|c|}
   \hline
   $\hspace{0.08cm} 1\hspace{0.08cm}$ & $\hspace{0.08cm} 2\hspace{0.08cm}$ \\
   \hline
   $\hspace{0.08cm} 3\hspace{0.08cm}$  \\
   \cline{1-1}
\end{tabular},
\begin{tabular}{|c|c|}
    \hline
    $\hspace{0.08cm} \overline{4}\hspace{0.08cm}$ & $\hspace{0.08cm} \overline{5}\hspace{0.08cm}$ \\
    \hline
    $\hspace{0.08cm} \overline{6}\hspace{0.08cm}$ \\ 
    \cline{1-1}
\end{tabular})$,\\

$F_3^B = (\begin{tabular}{|c|c|}
   \hline
   $\hspace{0.08cm} 1\hspace{0.08cm}$ & $\hspace{0.08cm} 3\hspace{0.08cm}$ \\
   \hline
   $\hspace{0.08cm} 2\hspace{0.08cm}$  \\
   \cline{1-1}
\end{tabular},
\begin{tabular}{|c|c|}
    \hline
    $\hspace{0.08cm} \overline{4}\hspace{0.08cm}$ & $\hspace{0.08cm} \overline{5}\hspace{0.08cm}$ \\
    \hline
    $\hspace{0.08cm} \overline{6}\hspace{0.08cm}$ \\ 
    \cline{1-1}
\end{tabular}),\
F_4^B = (\begin{tabular}{|c|c|}
   \hline
   $\hspace{0.08cm} 1\hspace{0.08cm}$ & $\hspace{0.08cm} 3\hspace{0.08cm}$ \\
   \hline
   $\hspace{0.08cm} 2\hspace{0.08cm}$  \\
   \cline{1-1}
\end{tabular},
\begin{tabular}{|c|c|}
    \hline
    $\hspace{0.08cm} \overline{4}\hspace{0.08cm}$ & $\hspace{0.08cm} \overline{6}\hspace{0.08cm}$ \\
    \hline
    $\hspace{0.08cm} \overline{5}\hspace{0.08cm}$ \\ 
    \cline{1-1}
\end{tabular}),$ \\

$CS_1^B = (\begin{tabular}{|c|c|}
   \hline
   $\hspace{0.08cm} 1\hspace{0.08cm}$ & $\hspace{0.08cm} 2\hspace{0.08cm}$ \\
   \hline
   $\hspace{0.08cm} 3\hspace{0.08cm}$  \\
   \cline{1-1}
\end{tabular},
\begin{tabular}{|c|c|}
    \hline
    $\hspace{0.08cm} \overline{4}\hspace{0.08cm}$ & $\hspace{0.08cm} \overline{5}\hspace{0.08cm}$ \\
    \hline
    $\hspace{0.08cm} \overline{6}\hspace{0.08cm}$ \\ 
    \cline{1-1}
\end{tabular}),\quad
CS_2^B = (\begin{tabular}{|c|c|}
   \hline
   $\hspace{0.08cm} 1\hspace{0.08cm}$ & $\hspace{0.08cm} 2\hspace{0.08cm}$ \\
   \hline
   $\hspace{0.08cm} 3\hspace{0.08cm}$  \\
   \cline{1-1}
\end{tabular},
\begin{tabular}{|c|c|}
    \hline
    $\hspace{0.08cm} \overline{4}\hspace{0.08cm}$ & $\hspace{0.08cm} \overline{6}\hspace{0.08cm}$ \\
    \hline
    $\hspace{0.08cm} \overline{5}\hspace{0.08cm}$ \\ 
    \cline{1-1}
\end{tabular})$,\\

$CS_3^B = (\begin{tabular}{|c|c|}
   \hline
   $\hspace{0.08cm} 1\hspace{0.08cm}$ & $\hspace{0.08cm} 3\hspace{0.08cm}$ \\
   \hline
   $\hspace{0.08cm} 2\hspace{0.08cm}$  \\
   \cline{1-1}
\end{tabular},
\begin{tabular}{|c|c|}
    \hline
    $\hspace{0.08cm} \overline{4}\hspace{0.08cm}$ & $\hspace{0.08cm} \overline{5}\hspace{0.08cm}$ \\
    \hline
    $\hspace{0.08cm} \overline{6}\hspace{0.08cm}$ \\ 
    \cline{1-1}
\end{tabular}),\quad
CS_4^B = (\begin{tabular}{|c|c|}
   \hline
   $\hspace{0.08cm} 1\hspace{0.08cm}$ & $\hspace{0.08cm} 3\hspace{0.08cm}$ \\
   \hline
   $\hspace{0.08cm} 2\hspace{0.08cm}$  \\
   \cline{1-1}
\end{tabular},
\begin{tabular}{|c|c|}
    \hline
    $\hspace{0.08cm} \overline{4}\hspace{0.08cm}$ & $\hspace{0.08cm} \overline{6}\hspace{0.08cm}$ \\
    \hline
    $\hspace{0.08cm} \overline{5}\hspace{0.08cm}$ \\ 
    \cline{1-1}
\end{tabular}).$

\section{Results}
\label{Results}

The nucleon–antinucleon system can have four possible $(I,S)$ combinations: (1,1),(1,0),(0,1)), and 
(0,0). However, our results show that the ground states  for $I=0$ and $I=1$ are identical. Therefore, we present only the results for $I=0$ in this section.

To investigate the possibility of nucleon–antinucleon bound state, we first calculate the expectation value of the Hamiltonian using the variational method while varying the distance between the two particles. Here, the distance $r_{N-\overline{N}}$ is defined as the relative separation between the centers of mass of the baryon and antibaryon as follows.
\begin{align}
    r_{N-\overline{N}} = \frac{1}{3}(\vec{r}_1 + \vec{r}_2 + \vec{r}_3 -\vec{r}_4 -\vec{r}_5 -\vec{r}_6).
\end{align}

The total energy as a function of the distance between a nucleon and an antinucleon for $S=0$ and $S=1$ is shown in Figs. \ref{mass(I=0,S=1)} and \ref{mass(I=0,S=0)}, respectively. Here, the threshold refers to the sum of the nucleon and antinucleon masses calculated within our model. These results indicate that the nucleon–antinucleon system has the possibility of forming a bound state for $S=0$.

\begin{figure}[H]
    \centering
    \includegraphics[width=1\linewidth]{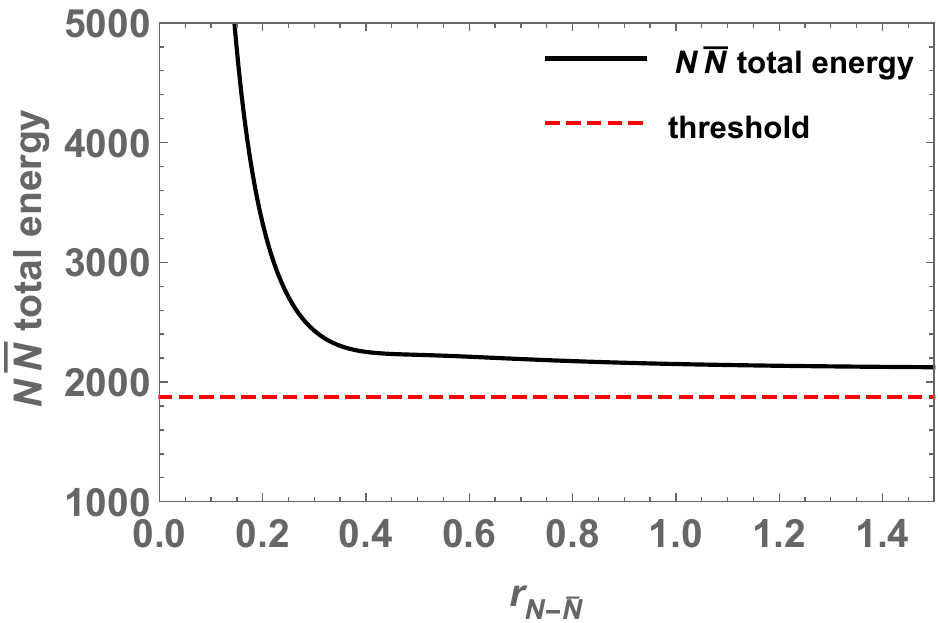}
    \caption{Nucleon-antinucleon total energy for $I=0$ and $S=1$. The red dashed line represents the threshold mass. The units on the $y-$ and $x-$ axes are MeV and fm, respectively.}
    \label{mass(I=0,S=1)}
\end{figure}

\begin{figure}[H]
    \centering
    \includegraphics[width=1\linewidth]{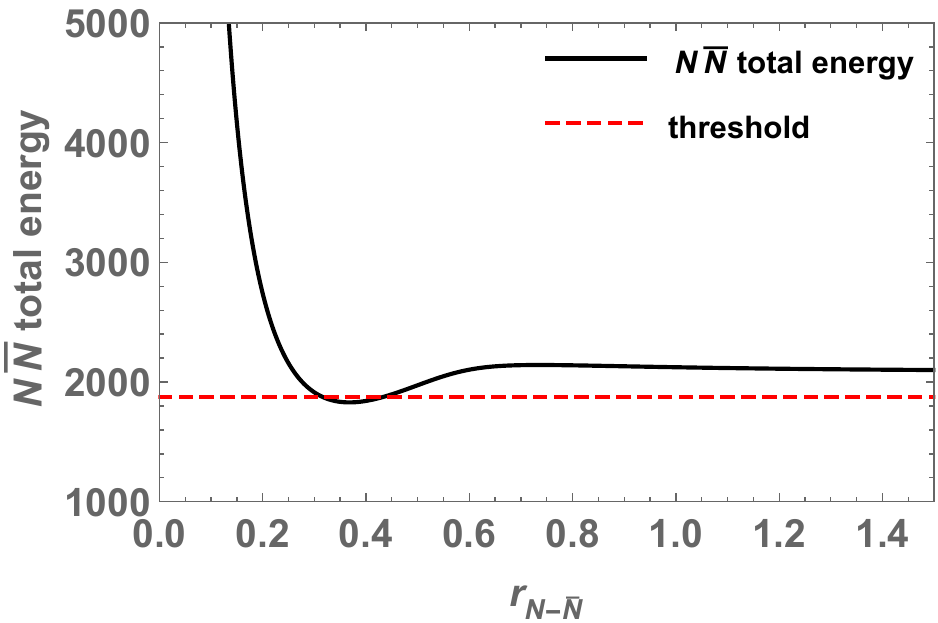}
    \caption{Nucleon-antinucleon total energy for $I=0$ and $S=0$. The red dashed line represents the threshold mass. The units are the same as in Fig. \ref{mass(I=0,S=1)}.}
    \label{mass(I=0,S=0)}
\end{figure}

To further investigate the interaction between the nucleon and the antinucleon, we extract only the interaction terms between them in the Hamiltonian. 
\begin{align}
    V_{N\overline{N}} =& H^C_{N\overline{N}}-H
    ^C_{N}-H^C_{\overline{N}} + H^{CS}_{N\overline{N}}-H^{CS}_{N}-H^{CS}_{\overline{N}} \nonumber \\
    & +H^3_{N\overline{N}}-H^3_{N}-H^3_{\overline{N}},
\end{align}
where $H^C$, $H^{CS}$, and $H^3$ are color-color, color-spin, and three-quark potential parts of the Hamiltonian, respectively.
The results are shown in Figs. \ref{interaction(I=0,S=1)},and  \ref{interaction(I=0,S=0)} for $S=1$ and $S=0$, respectively. In this study, we also examine the effect of the three-quark potentials on the nucleon–antinucleon interaction by plotting results with and without their inclusion. As shown in the figure, the three-quark potential is repulsive in both the $S=0$ and 
$S=1$ cases. However, the two-quark interaction is significantly more attractive, so that the total interactions show large attraction at short distance for both cases.

\begin{figure}[H]
    \centering
    \includegraphics[width=1\linewidth]{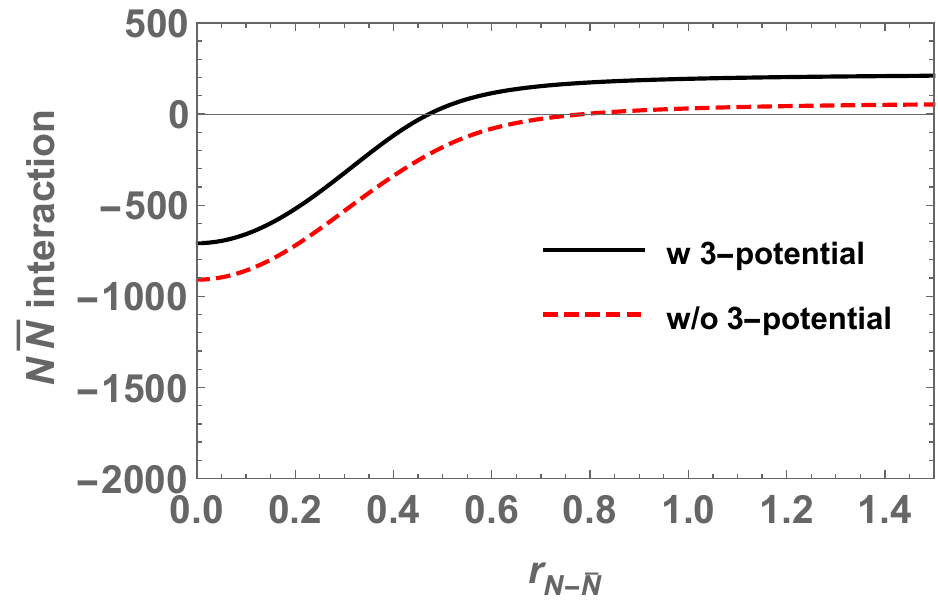}
    \caption{Nucleon-antinucleon interaction for $I=0$ and $S=1$. The red dashed line represents the interaction without three-quark potentials. The units are the same as in Fig. \ref{mass(I=0,S=1)}.}
    \label{interaction(I=0,S=1)}
\end{figure}

\begin{figure}[H]
    \centering
    \includegraphics[width=1\linewidth]{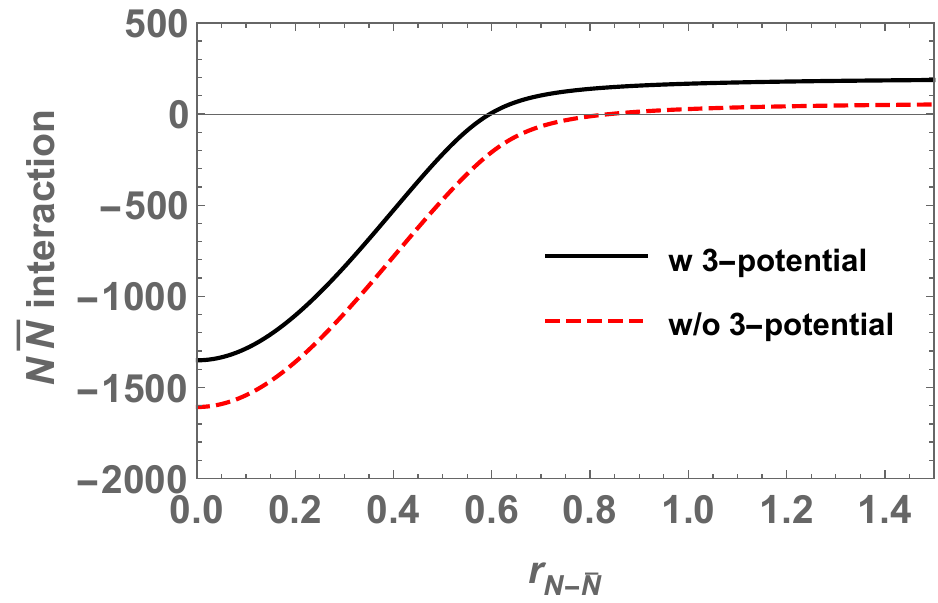}
    \caption{Nucleon-antinucleon interaction for $I=0$ and $S=0$. The red dashed line represents the interaction without three-quark potentials. The units are the same as in Fig. \ref{mass(I=0,S=1)}.}
    \label{interaction(I=0,S=0)}
\end{figure}

Additionally, to investigate which color basis is dominant in the ground state, we present the color probabilities as a function of distance between the nucleon and the antinucleon in Fig. \ref{color(I=0,S=1)} and \ref{color(I=0,S=0)} for $S=1$ and $S=0$, respectively. As shown in the figures, the octet–octet color component is dominant at short range in both the $S=0,1$ cases. 
It should be noted that, in the 
$S=0$ case, the system becomes most attractive at a separation of approximately 0.4 fm. At this distance, the octet–octet and decuplet–antidecuplet color bases for $S=0$ have higher probabilities than in the 
$S=1$ case.

In Table \ref{mass-table}, We represent the mass of the nucleon–antinucleon system, the expectation values of each potential, and the variational parameters. The ground state wave functions are as follows.

\begin{itemize}
    \item $(I=0,S=1)$: $\psi_3^{(0,1)}$.
    \item $(I=0,S=0)$: $0.33\psi_3^{(0,0)}+0.63\psi_4^{(0,0)} -0.57\psi_5^{(0,0)} +0.41 \psi_6^{(0,0)}$.
\end{itemize}
Here, the superscripts represent the isospin and spin quantum numbers. The explicit forms of the wave functions corresponding to each isospin and spin are presented in the Appendix \ref{wave-function}.
For $S=1$ case, the color wave function of ground state is color singlet-singlet which is $C_1$. However, for $S=0$ case, color octet-octet has more probability than color singlet-singlet. Therefore, we conclude that this octet-octet dominance can be regarded as an indication of the bound state possibility of nucleon-antinucleon with $S=0$.

An additional important point to consider is that in the nucleon–antinucleon system, the proton–antiproton pair experiences an electric attraction, unlike the repulsion between two protons. At a separation of 0.4 fm, the electric potential between the proton and antiproton is approximately –3.6 MeV. Although this attractive potential is not sufficient to bind the system in the $S = 1$ state, it can enhance the binding energy in the $S = 0$ state, potentially making the bound state more favorable.

\begin{figure}[H]
    \centering
    \includegraphics[width=1\linewidth]{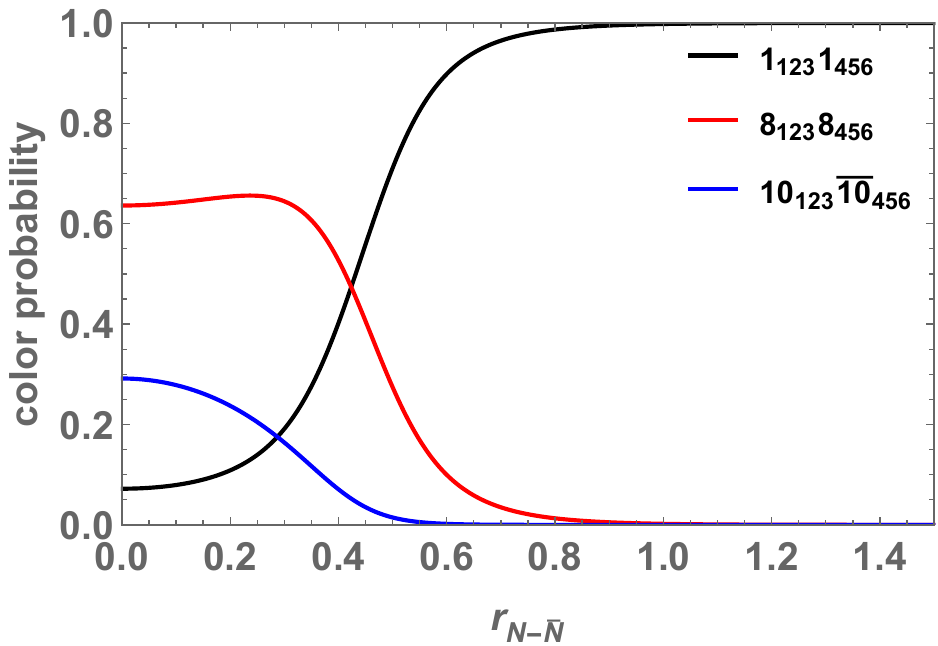}
    \caption{Color probability of nucleon-antinucleon for $I=0$ and $S=1$.}
    \label{color(I=0,S=1)}
\end{figure}

\begin{figure}[H]
    \centering
    \includegraphics[width=1\linewidth]{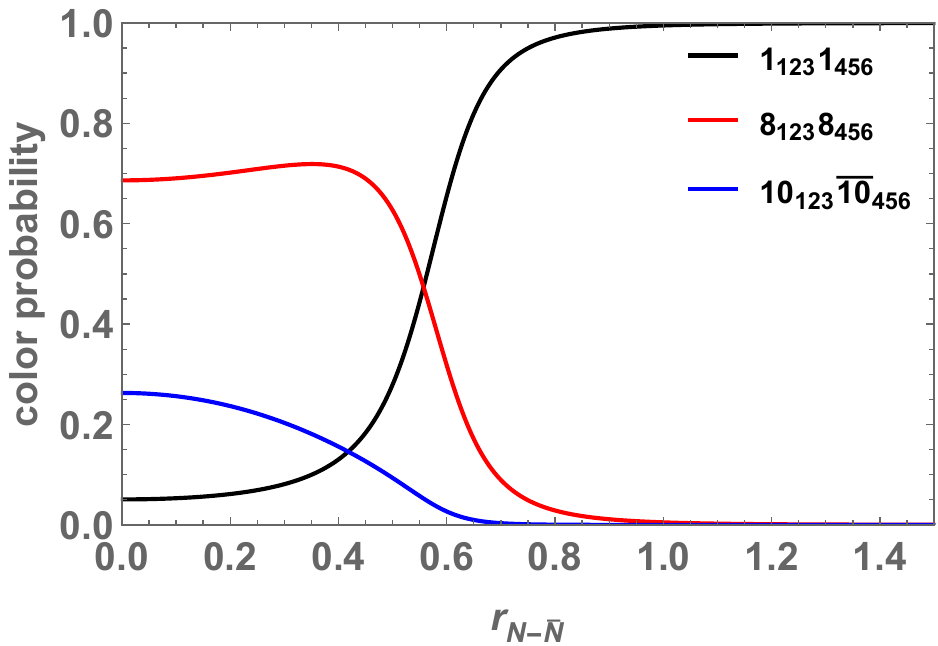}
    \caption{Color probability of nucleon-antinucleon for $I=0$ and $S=0$.}
    \label{color(I=0,S=0)}
\end{figure}

\begin{widetext}
\begin{table*}[ht]
\centering
\begin{tabular}{c|c|c|c|c|c|c|c|c|c}
\hline
\hline	
Nucleon-antinucleon & Mass & $\langle H_C \rangle$ & $\langle H_{CS}\rangle$ & $\langle L^{C-C}\rangle$ & $\langle L^{S-S}\rangle$ & $\langle L^{C-S}\rangle$ & $a_1$ & $a_2$ & $a_3$ \\
\hline
$(I=0,1,S=1)$ & 2099.9 & -1236.33 & -319.24 & -46.39 & -4.58 & 73.32 & 2.3 & 2.4 & 0.001 \\
\hline
$(I=0,1,S=0)$ & 1828.88 & -1356.65 & -1177.46 & 138.27 & -16.81 & -2.29 & 2.4 & 2.4 & 3.1 \\
\hline
\hline
\end{tabular}
\caption{Results for nucleon-antinucleon. All expectations values are calculated with the ground state. Units are MeV and $\text{fm}^{-2}$ for the energies and the variational parameters, respectively.}
\label{mass-table}
\end{table*}
\end{widetext}

\section{Summary}
\label{Summary}

In this work, we study the nucleon-antinucleon configuration using a constituent quark model. We first construct the nucleon-antinucleon wave function satisfying the Pauli principle for each multiquark and anti-multiquark system. We investigate the possibility of a bound state of nucleon-antinucleon using a Hamiltonian including both color-color and color-spin interactions. Our results show that there is strong possibility of bound state for $(I,S)=(0,0),(1,0)$ cases.

We should note that instead of using the nucleon-antinucleon basis, we may equivalently use the three-meson configuration. In such basis, the asymptotic configuration will be the three separated pions.  On the other hand, when the three pions are close together, a large amount of additional kinetic energy and inter-meson interactions will lead to the ground state basis that we have obtained in this work, which could be also expressed in terms of an appropriate color basis and Jacobi coordinates in the three-meson basis.

\appendix

\section{Color-color matrices}
\label{Color-color matrices}

Here, we represent the color-color matrices for baryon-antibaryon configuration. The color basis set of these matrices are represented in Sec.\ref{Color basis}.

$\lambda^c_1 \lambda^c_2 = \left(
    \begin{array}{cccccc}
 -\frac{8}{3} & 0 & 0 & 0 & 0 & 0 \\
 0 & \frac{4}{3} & 0 & 0 & 0 & 0 \\
 0 & 0 & \frac{4}{3} & 0 & 0 & 0 \\
 0 & 0 & 0 & -\frac{8}{3} & 0 & 0 \\
 0 & 0 & 0 & 0 & -\frac{8}{3} & 0 \\
 0 & 0 & 0 & 0 & 0 & \frac{4}{3} \\
    \end{array}
    \right)$,\\
    
$\lambda^c_1 \lambda^c_4 = \left(
\begin{array}{cccccc}
 0 & -\sqrt{2} & \sqrt{\frac{2}{3}} & \sqrt{\frac{2}{3}} & -\frac{\sqrt{2}}{3} & 0
   \\
 -\sqrt{2} & -\frac{5}{2} & \frac{5}{2 \sqrt{3}} & \frac{5}{2 \sqrt{3}} &
   -\frac{3}{2} & -\frac{\sqrt{5}}{3} \\
 \sqrt{\frac{2}{3}} & \frac{5}{2 \sqrt{3}} & -\frac{5}{6} & -\frac{3}{2} &
   -\frac{1}{2 \sqrt{3}} & -\sqrt{\frac{5}{3}} \\
 \sqrt{\frac{2}{3}} & \frac{5}{2 \sqrt{3}} & -\frac{3}{2} & -\frac{5}{6} &
   -\frac{1}{2 \sqrt{3}} & -\sqrt{\frac{5}{3}} \\
 -\frac{\sqrt{2}}{3} & -\frac{3}{2} & -\frac{1}{2 \sqrt{3}} & -\frac{1}{2 \sqrt{3}}
   & -\frac{7}{6} & -\sqrt{5} \\
 0 & -\frac{\sqrt{5}}{3} & -\sqrt{\frac{5}{3}} & -\sqrt{\frac{5}{3}} & -\sqrt{5} &
   -\frac{8}{3} \\
\end{array}
\right)$,\\

$\lambda^c_4 \lambda^c_5 = \left(
\begin{array}{cccccc}
 -\frac{8}{3} & 0 & 0 & 0 & 0 & 0 \\
 0 & \frac{4}{3} & 0 & 0 & 0 & 0 \\
 0 & 0 & -\frac{8}{3} & 0 & 0 & 0 \\
 0 & 0 & 0 & \frac{4}{3} & 0 & 0 \\
 0 & 0 & 0 & 0 & -\frac{8}{3} & 0 \\
 0 & 0 & 0 & 0 & 0 & \frac{4}{3} \\
\end{array}
\right)$.\\

We can obtain the remaining color-color matrices using the permutation symmetry of the nucleon-antinucleon configuration. Additionally, we can check that the color-color matrices satisfy the following condition.
\begin{align}
  \sum_{i<j}^6 \lambda^c_i \lambda^c_j = \left(
\begin{array}{cccccc}
 -16 & 0 & 0 & 0 & 0 & 0 \\
 0 & -16 & 0 & 0 & 0 & 0 \\
 0 & 0 & -16 & 0 & 0 & 0 \\
 0 & 0 & 0 & -16 & 0 & 0 \\
 0 & 0 & 0 & 0 & -16 & 0 \\
 0 & 0 & 0 & 0 & 0 & -16 \\
\end{array}
\right) \nonumber  
\end{align}

\section{Wave function of $q^3 \overline{q}^3$}
\label{wave-function}

\subsubsection{$(I=1,S=1)$}
\begin{align}
    \psi_1^{(1,1)} &= F_1^A \otimes C_1 \otimes [(12)_{S=1}3]_{S=\frac{3}{2}}[(45)_{S=1}6]_{S=\frac{3}{2}} \\
    \psi_2^{(1,1)} &= F_1^A \otimes (C_2 \sim C_5) \otimes [123]_{S=\frac{1}{2}}[456]_{S=\frac{1}{2}} \\
    \psi_3^{(1,1)} &= \frac{1}{\sqrt{2}} \{ F_1^B \otimes C_1 \otimes [(12)_{S=1}3]_{S=\frac{3}{2}}[(45)_{S=1}6]_{S=\frac{1}{2}} \nonumber \\
    & \ - F_2^B \otimes C_1 \otimes [(12)_{S=1}3]_{S=\frac{3}{2}}[(45)_{S=0}6]_{S=\frac{1}{2}} \} \\
    \psi_4^{(1,1)} &= \frac{1}{2} \{ F_1^B \otimes (C_3 \otimes [(12)_{S=0}3]_{S=\frac{1}{2}}[(45)_{S=1}6]_{S=\frac{3}{2}} \nonumber \\
    & \ -C_5 \otimes [(12)_{S=1}3]_{S=\frac{1}{2}}[(45)_{S=1}6]_{S=\frac{3}{2}}) \nonumber \\
    & \ - F_2^B \otimes (C_2 \otimes [(12)_{S=0}3]_{S=\frac{1}{2}}[(45)_{S=1}6]_{S=\frac{3}{2}} \nonumber \\
    & \ -C_4 \otimes [(12)_{S=1}3]_{S=\frac{1}{2}}[(45)_{S=1}6]_{S=\frac{3}{2}}) \} \\
    \psi_5^{(1,1)} &= F_1^B \otimes (C_2 \sim C_5) \otimes [123]_{S=\frac{1}{2}}[456]_{S=\frac{1}{2}} \\
    \psi_6^{(1,1)} &= \frac{1}{\sqrt{2}} \{ -F_1^C \otimes C_1 \otimes [(12)_{S=1}3]_{S=\frac{1}{2}}[(45)_{S=1}6]_{S=\frac{3}{2}} \nonumber \\
    & \ - F_2^C \otimes C_1 \otimes [(12)_{S=0}3]_{S=\frac{1}{2}}[(45)_{S=1}6]_{S=\frac{3}{2}} \} \\
    \psi_7^{(1,1)} &= \frac{1}{2} \{ F_1^C \otimes (C_4 \otimes [(12)_{S=1}3]_{S=\frac{3}{2}}[(45)_{S=0}6]_{S=\frac{1}{2}} \nonumber \\
    & \ -C_5 \otimes [(12)_{S=1}3]_{S=\frac{3}{2}}[(45)_{S=1}6]_{S=\frac{1}{2}}) \nonumber \\
    & \ - F_2^C \otimes (C_2 \otimes [(12)_{S=1}3]_{S=\frac{3}{2}}[(45)_{S=0}6]_{S=\frac{1}{2}} \nonumber \\
    & \ -C_3 \otimes [(12)_{S=1}3]_{S=\frac{3}{2}}[(45)_{S=1}6]_{S=\frac{1}{2}}) \}
\end{align}
\begin{align}
    \psi_8^{(1,1)} &= (F_1^C \sim F_2^C) \otimes (C_2 \sim C_5) \otimes [123]_{S=\frac{1}{2}}[456]_{S=\frac{1}{2}} \\
    \psi_9^{(1,1)} &= (F_1^D \sim F_4^D) \otimes C_1 \otimes [123]_{S=\frac{1}{2}}[456]_{S=\frac{1}{2}} \\
    \psi_{10}^{(1,1)} &= (F_1^D \sim F_4^D) \otimes (C_2 \sim C_5) \otimes [123]_{S=\frac{3}{2}}[456]_{S=\frac{3}{2}} \\
    \psi_{11}^{(1,1)} &= (F_1^D \sim F_4^D) \otimes (C_2 \sim C_5) \otimes [123]_{S=\frac{3}{2}}[456]_{S=\frac{1}{2}} \\
    \psi_{12}^{(1,1)} &= (F_1^D \sim F_4^D) \otimes (C_2 \sim C_5) \otimes [123]_{S=\frac{1}{2}}[456]_{S=\frac{3}{2}} \\
    \psi_{13}^{(1,1)} &= (F_1^D \sim F_4^D) \otimes (C_2 \sim C_5) \otimes [123]_{S=\frac{1}{2}}[456]_{S=\frac{1}{2}} \\
    \psi_{14}^{(1,1)} &= (F_1^D \sim F_4^D) \otimes C_6 \otimes [123]_{S=\frac{1}{2}}[456]_{S=\frac{1}{2}} 
\end{align}

\subsubsection{$(I=1,S=0)$}
\begin{align}
    \psi_1^{(1,0)} &= F_1^A \otimes C_1 \otimes [(12)_{S=1}3]_{S=\frac{3}{2}}[(45)_{S=1}6]_{S=\frac{3}{2}} \\
    \psi_2^{(1,0)} &= F_1^A \otimes (C_2 \sim C_5) \otimes [123]_{S=\frac{1}{2}}[456]_{S=\frac{1}{2}} \\
    \psi_3^{(1,0)} &= F_1^B \otimes (C_2 \sim C_5) \otimes [123]_{S=\frac{1}{2}}[456]_{S=\frac{1}{2}} \\
    \psi_4^{(1,0)} &= F_1^C \otimes (C_2 \sim C_5) \otimes [123]_{S=\frac{1}{2}}[456]_{S=\frac{1}{2}} \\
    \psi_5^{(1,0)} &= (F_1^D \sim F_4^D) \otimes C_1 \otimes [123]_{S=\frac{1}{2}}[456]_{S=\frac{1}{2}} \\
    \psi_6^{(1,0)} &= (F_1^D \sim F_4^D) \otimes (C_2 \sim C_5) \otimes [123]_{S=\frac{3}{2}}[456]_{S=\frac{3}{2}} \\
    \psi_7^{(1,0)} &= (F_1^D \sim F_4^D) \otimes (C_2 \sim C_5) \otimes [123]_{S=\frac{1}{2}}[456]_{S=\frac{1}{2}} \\
    \psi_8^{(1,0)} &= (F_1^D \sim F_4^D) \otimes C_6 \otimes [123]_{S=\frac{1}{2}}[456]_{S=\frac{1}{2}} 
\end{align}

\subsubsection{$(I=0,S=1)$}
\begin{align}
    \psi_1^{(0,1)} &= F_1^A \otimes C_1 \otimes [(12)_{S=1}3]_{S=\frac{3}{2}}[(45)_{S=1}6]_{S=\frac{3}{2}} \\
    \psi_2^{(0,1)} &= F_1^A \otimes (C_2 \sim C_5) \otimes [123]_{S=\frac{1}{2}}[456]_{S=\frac{1}{2}} \\
    \psi_3^{(0,1)} &= (F_1^B \sim F_4^B) \otimes C_1 \otimes [123]_{S=\frac{1}{2}}[456]_{S=\frac{1}{2}} \\
    \psi_4^{(0,1)} &= (F_1^B \sim F_4^B) \otimes (C_2 \sim C_5) \otimes [123]_{S=\frac{3}{2}}[456]_{S=\frac{3}{2}} \\
    \psi_5^{(0,1)} &= (F_1^B \sim F_4^B) \otimes (C_2 \sim C_5) \otimes [123]_{S=\frac{3}{2}}[456]_{S=\frac{1}{2}} \\
    \psi_6^{(0,1)} &= (F_1^B \sim F_4^B) \otimes (C_2 \sim C_5) \otimes [123]_{S=\frac{1}{2}}[456]_{S=\frac{3}{2}} \\
    \psi_7^{(0,1)} &= (F_1^B \sim F_4^B) \otimes (C_2 \sim C_5) \otimes [123]_{S=\frac{1}{2}}[456]_{S=\frac{1}{2}} \\
    \psi_8^{(0,1)} &= (F_1^B \sim F_4^B) \otimes C_6 \otimes [123]_{S=\frac{1}{2}}[456]_{S=\frac{1}{2}} 
\end{align}

\subsubsection{$(I=0,S=0)$}
\begin{align}
    \psi_1^{(0,0)} &= F_1^A \otimes C_1 \otimes [(12)_{S=1}3]_{S=\frac{3}{2}}[(45)_{S=1}6]_{S=\frac{3}{2}} \\
    \psi_2^{(0,0)} &= F_1^A \otimes (C_2 \sim C_5) \otimes [123]_{S=\frac{1}{2}}[456]_{S=\frac{1}{2}} \\
    \psi_3^{(0,0)} &= (F_1^B \sim F_4^B) \otimes C_1 \otimes [123]_{S=\frac{1}{2}}[456]_{S=\frac{1}{2}} \\
    \psi_4^{(0,0)} &= (F_1^B \sim F_4^B) \otimes (C_2 \sim C_5) \otimes [123]_{S=\frac{3}{2}}[456]_{S=\frac{3}{2}} \\
    \psi_5^{(0,0)} &= (F_1^B \sim F_4^B) \otimes (C_2 \sim C_5) \otimes [123]_{S=\frac{1}{2}}[456]_{S=\frac{1}{2}} \\
    \psi_6^{(0,0)} &= (F_1^B \sim F_4^B) \otimes C_6 \otimes [123]_{S=\frac{1}{2}}[456]_{S=\frac{1}{2}} 
\end{align}

\section*{Acknowledgments}
This work was supported by the Korea National Research Foundation under the grant No. 2023R1A2C300302312, No.
RS-2023-00280831 and No. RS-2025-23963552.

\end{document}